\documentclass[fleqn, 12pt] {article}
\usepackage{epsf}
\begin{document}
\title{Spiky development at the interface in Rayleigh-Taylor instability: Layzer approximation with second harmonic }
\author{ M. R. Gupta\thanks{e-mail: mrgupta$_{-}$cps@yahoo.co.in}, Rahul Banerjee\thanks{e-mail: rbanerjee.math@gmail.com}, Labakanta Mandal, S. Roy, Manoranjan Khan\thanks{e-mail: mkhan$_{-}$ju@yahoo.com} \\
Department of Instrumentation Science \& Centre for Plasma Studies
\\Jadavpur University, Kolkata-700032, India\\}
\date{}

\maketitle
\begin{abstract}
Layzer's approximation method for investigation of two fluid
interface structures associated with Rayleigh Taylor instability
for arbitrary Atwood number is extended with the inclusion of
second harmonic mode leaving out the zeroth harmonic one. The
modification makes the fluid velocities vanish at infinity and
leads to avoidance of the need to make the unphysical assumption
of the existence of a time dependent source at infinity.

The present analysis shows that for an initial interface
perturbation with curvature exceeding $1/(2\sqrt{A})$, where $A$
is the Atwood number there occurs an almost free fall of the spike
with continuously increasing sharpening as it falls. The curvature
at the tip of the spike also increases with Atwood number. Certain
initial condition may also result in occurrence of finite time
singularity as found in case of conformal mapping technique used
earlier. However bubble growth rate is not appreciably affected.

\end{abstract}
\newpage

Hydrodynamic instabilities such as Rayleigh Taylor
instability(RTI) which sets in when a lighter fluid supports a
heavier fluid against gravity or Richtmyer Meshkov
instability(RMI) which is initiated when a shock passes an
interface between two fluids with different acoustic impedances
are of increasing importance in a wide range of physical phenomena
starting from inertial confinement fusion (ICF) to astrophysical
ones like supernova explosions. In ICF, the capsule shell
undergoes the RTI both in the acceleration and deceleration
phases. RTI can retard the formation of the hot spot by the cold
RTI spike of capsule shell resulting in the destruction of the
ignition hot spot or autoignition${\cite{lin04}-\cite{at04}}$. The
hydrodynamic instabilities lead to development of heavy fluid
'spikes' penetrating into the lighter fluid and 'bubbles' of
lighter fluid rising through the heavier fluid. Different
approaches have been used for the study of such problems. Among
these Layzer's${\cite{dl55}}$ approach applied to single mode
potential flow model${\cite{jh94}-\cite{ss03}}$ is a useful one
giving approximate estimate of both Rayleigh Taylor and Richtmyer
Meshkov instability evolution. The bubbles were shown by
Zhang${\cite{qz98}}$ to rise at a rate tending asymptotically to a
terminally constant velocity while spikes were shown to descend
with a constant acceleration. However, whether for bubbles or for
the spikes, Zhang's analysis was applicable only for Atwood number
$A=1$, i.e, only for fluid-vacuum interface. An extension to
arbitrary value of Atwood number $A$ was done by
Goncharov${\cite{vg02}}$. Within limitations of Layzer's model as
pointed out by Mikaelian${\cite{mik08}}$ bubbles were shown to
rise with a velocity tending to an asymptotic value dependent on
$A$ and having a fairly close agreement with the simulation
results of Ramaprabhu et al.${\cite{pr06}}$. But the spikes were
found to descend with a terminal constant velocity in contrast to
a constant acceleration as obtained by Zhang${\cite{qz98}}$ for
$A=1$.

Asymptotic spike evolution in Rayleigh Taylor instability behaving
almost as a free fall was obtained by Clavin and
Williams${\cite{cl05}}$ and also by Duchemin et al.${\cite{du05}}$
by conformal mapping method. Associated with the free fall of the
spike, the surface curvature of the spike was also found to
increase with time (i.e the spike sharpens as it falls).

In the single mode Layzer model with generalization
${\cite{vg02}}$ for arbitrary Atwood number the equation to the
interface  taken in X-Y plane as
\begin{eqnarray}\label{eq:1}
y\equiv\eta(x,t)=\eta_{0}(t)+\eta_{2}(t)x^2;
\end{eqnarray}
with $\eta_{0}(t)>0$, $\eta_{2}(t)<0$ for bubble while
$\eta_{0}(t)<0$, $\eta_{2}(t)>0$ for spike. The velocity potential
describing the motion of the heavier fluid (density $\rho_{h}$)
and the lighter fluid (density $\rho_{l}$) are ( gravity $g$ is
along the negative $y$ direction)
\begin{eqnarray}\label{eq:2}
\phi_h(x,y,t)=a(t)\cos{(kx)}e^{-k(y-\eta_0(t))}; \quad y>0
\quad(heavier fluid)
\end{eqnarray}
\begin{eqnarray}\label{eq:3}
\phi_l(x,y,t)=b_{0}(t)y+b_1(t)\cos{(kx)}e^{k(y-\eta_0(t))}; \quad
y<0 \quad (lighter fluid)
\end{eqnarray}
where $k$ is the wave number and $a(t),b_{0}(t),b_1(t)$ are
amplitudes.This conventional single mode Layzer model has the
drawback that rather than conforming to the physical requirement:
$v_{ly}\rightarrow 0$ as $y\rightarrow-\infty$ it necessitates the
assumption of a time dependent source at
$y\rightarrow-\infty$${\cite{ab03}}$. To avoid this difficulty we
modify the single mode Layzer model by replacing the zeroth mode
term $b_{0}(t)y$ in Eq.(3) by a second harmonic term viz,
\begin{eqnarray}\label{eq:4}
\phi_l(x,y,t)=b_1(t)\cos{(kx)}e^{k(y-\eta_0(t))}+b_2(t)\cos{(2kx)}e^{2k(y-\eta_0(t))};
\quad y<0
\end{eqnarray}
 Eqs.(2) and (4) give
\begin{eqnarray}\label{eq:5}
v_{hy}(y\rightarrow\infty)=-\frac{\partial\phi_{h}}{\partial
y}]_{y\rightarrow\infty}=0
\end{eqnarray}
\begin{eqnarray}\label{eq:6}
v_{ly}(y\rightarrow-\infty)=-\frac{\partial\phi_{l}}{\partial
y}]_{y\rightarrow-\infty}=0
\end{eqnarray}
The kinematic boundary conditions at the interface (1) are
\begin{eqnarray}\label{eq:7}
\frac{\partial\eta}{\partial t}+v_{hx}\frac{\partial\eta}{\partial
x}=v_{hy}
\end{eqnarray}
\begin{eqnarray}\label{eq:8}
\frac{\partial\eta}{\partial x}(v_{hx}-v_{lx})=v_{hy}-v_{ly}
\end{eqnarray}
Setting the pressure boundary condition $p_{h}=p_{l}$ in
Bernoulli's equation for the heavier and lighter fluids leads
to${\cite{qz98}\cite{vg02}\cite{ss03}\cite{ss07}-\cite{rb11}}$
\begin{eqnarray}\label{eq:9}
\rho_{h}[-\frac{\partial \phi_{h}}{\partial t}+
\frac{1}{2}(\vec{\nabla} \phi_{h})^{2}+ g
\eta]-\rho_{l}[-\frac{\partial \phi_{l}}{\partial t}+
\frac{1}{2}(\vec{\nabla} \phi_{l})^{2}+ g \eta]=f_{h}(t)-f_{l}(t)
\end{eqnarray}
Following the usual procedure${\cite{mrg09}-\cite{rb11}}$ ,i.e,
expanding $\eta(x,t)$ and the velocity potentials in powers of
$(kx)$ and equating coefficients of $(kx)^{r},(r=0,2)$ we obtain
from Eqs.(7)-(9) the evolution equation for the RT bubbles/spikes
(nondimensionalized) tip elevation $\xi_{1}=k\eta_{0}$, curvature
$\xi_{2}=\eta_{2}/k$ and velocity $\xi_{3}=k^2 a/\sqrt{kg}$
\begin{eqnarray}\label{eq:10}
\frac{d\xi_1}{d\tau}=\xi_{3}
\end{eqnarray}
\begin{eqnarray}\label{eq:11}
\frac{d\xi_2}{d\tau}=-(3\xi_2 + \frac{1}{2})\xi_{3}
\end{eqnarray}
\begin{eqnarray}\label{eq:12}
\frac{d\xi_{3}}{d\tau}=-\frac{[2\xi_{2}(\xi_{2}-\frac{1}{2})^2+\xi_{3}^2(\xi_{2}-\alpha)(\xi_{2}-\beta)]}{[2(\xi_{2}-\frac{1}{2})(\xi_{2}^2
-\frac{1}{4}\frac{r+1}{r-1})]}
\end{eqnarray}
where
\begin{eqnarray}\label{eq:13}
\alpha,\beta=\frac{(r+4)\pm\sqrt{12r+13}}{2(r-1)};
r=\frac{\rho_{h}}{\rho_{l}}
\end{eqnarray}
and
\begin{eqnarray}\label{eq:14}
\tau=t\sqrt{kg}
\end{eqnarray}
is nondimensionalized time.

Starting from a set of initial values $\xi_{1}>0$, $\xi_{2}<0$ and
$\xi_{3}>0$ which correspond to the description of temporal
evolution of the tip of the bubble we arrive at the asymptotic
value ($\tau\rightarrow\infty$)
\begin{eqnarray}\label{eq:15}
\xi_{2}\rightarrow-\frac{1}{6}
\end{eqnarray}
and
\begin{eqnarray}\label{eq:16}
[\xi_{3}]_{asymp}=\sqrt{\frac{8A}{3(5A+3)}}>
[\xi_{3}]_{asymp}^{classical}=\sqrt{\frac{2}{3}\frac{A}{1+A}}
\end{eqnarray}
(by classical we mean the single mode Layzer approximation as used
by Goncharov${\cite{vg02}}$). Two values coincide as
$A(=\frac{\rho_{h}-\rho_{l}}{\rho_{h}+\rho_{l}})\rightarrow1$. The
growth rate of the development of the height of the bubble tip is
shown in Figure 1 and compared with classical value. It is seen
that presence or absence of a source does not give rise to any
qualitatively significant change in the growth rate of the bubble
height${\cite{vg02}\cite{ss01}\cite{ss07}}$.

To get spike like behavior of the perturbation of the interface we
used
\begin{eqnarray}\label{eq:17}
\xi_{1}<0,\quad \xi_{2}>0 \quad and \quad \xi_{3}<0
\end{eqnarray}
 Corresponding to a start from an initial value
 \begin{eqnarray}\label{eq:18}
[\xi_{2}]_{initial}>\frac{1}{2}\sqrt{\frac{r+1}{r-1}},\quad
[\xi_{3}]_{initial}<0
\end{eqnarray}
 Eq.(11) shows that $\xi_{2}$ increases
monotonically ($\frac{d\xi_{2}}{d\tau}>0$ for all $\tau>0$) while
from Eq.(12) it follows that the depth of the spike tip below the
surface of separation increases continuously ($\frac{d\xi_{1}}{d
\tau }=\xi_{3}<0$ and $\frac{d\xi_{3}}{d\tau}<0$). These are shown
in Figures 2(a) and 2(b) by plotting $\xi_{2}(\tau)$ and
$\xi_{3}(\tau)$ as function of $\tau$ obtained from numerical
solution of Eqs.(11) and (12) by employing fifth order
Runge-Kutta-Fehlberg method. The initial value taken are
$[\xi_{2}]_{initial}=1.0$ and $[\xi_{3}]_{initial}=-0.5$ which
satisfy condition (17) for all the following three values of $r=2
(A=\frac{1}{3})$, $r=5 (A=\frac{2}{3})$, $r=20 (A=\frac{19}{21})$.
The value of $\xi_{2}$ which represents the curvature at the tip
of the spike is an increasing function of $\tau$ for every value
of the Atwood number $A$ (Figure 2(a)). Moreover for every given
value of $\tau$ the curvature $\xi_{2}$ increases with $A$. This
implies that the spike continues to sharpen with time as well as
with increasing Atwood number and is explicitly shown in Figure 3.
Figure 2(b) shows that except very close to the starting instant
the spike descends with a constant acceleration $\simeq -g$ (i.e,
nearly a free fall). This agrees with the
conclusions${\cite{qz98}}$ for Atwood number $A=1$.

The time development of spiky behavior for
$[\xi_{2}]_{initial}>\frac{1}{2}\sqrt{\frac{r+1}{r-1}}$ and
$[\xi_{3}]_{initial}<0$ is demonstrated in Figures4(a) and 4(b).
This is shown both for increasing $A$ with fixed $\tau$ (Figure
4(a)) and with increasing $\tau$ for given $A$ (Figure 4(b)). But
for
$\frac{1}{2}<[\xi_{2}]_{initial}<\frac{1}{2}\sqrt{\frac{r+1}{r-1}}$
with $[\xi_{3}]_{initial}<0$ one encounters development of finite
time singularity i.e, $\xi_{2}\rightarrow\infty$ and
$\xi_{3}\rightarrow-\infty$ at a finite value of $\tau$. The
possibility of the occurrence of such an eventuality at (or near)
the tip of the spike is also found to arise when the RT
instability is addressed by conformal mapping
method${\cite{tb03}-\cite{tn93}}$ as mentioned by Clavin and
Williams${\cite{cl05}}$.

Finally for a trajectory starting from $(\xi_{3})_{initial}<0$ and
$0<(\xi_{2})_{initial}<\frac{1}{2}\sqrt{\frac{r+1}{r-1}}=\frac{1}{2}\sqrt{\frac{1}{A}}$
one finds that $\xi_{2}$ continues to increase towards
$\xi_{2}=\frac{1}{2}$, i.e, the spike continues to sharpen as time
progresses and its speed of fall slowly decreases in magnitude.
Because of the presence of singularity at
$\xi_{2}=\frac{1}{2}$(Eq.(12)), it is not possible to continue the
numerical integration towards and beyond this point. This is shown
in Figure 5 for initial values in the domain mentioned above.

\section*{ACKNOWLEDGMENTS }
This work is supported by the C.S.I.R, Government of India under
ref. no. R-10/B/1/09.

\begin{figure}[p]
\vbox{\hskip 1.cm \epsfxsize=12cm \epsfbox{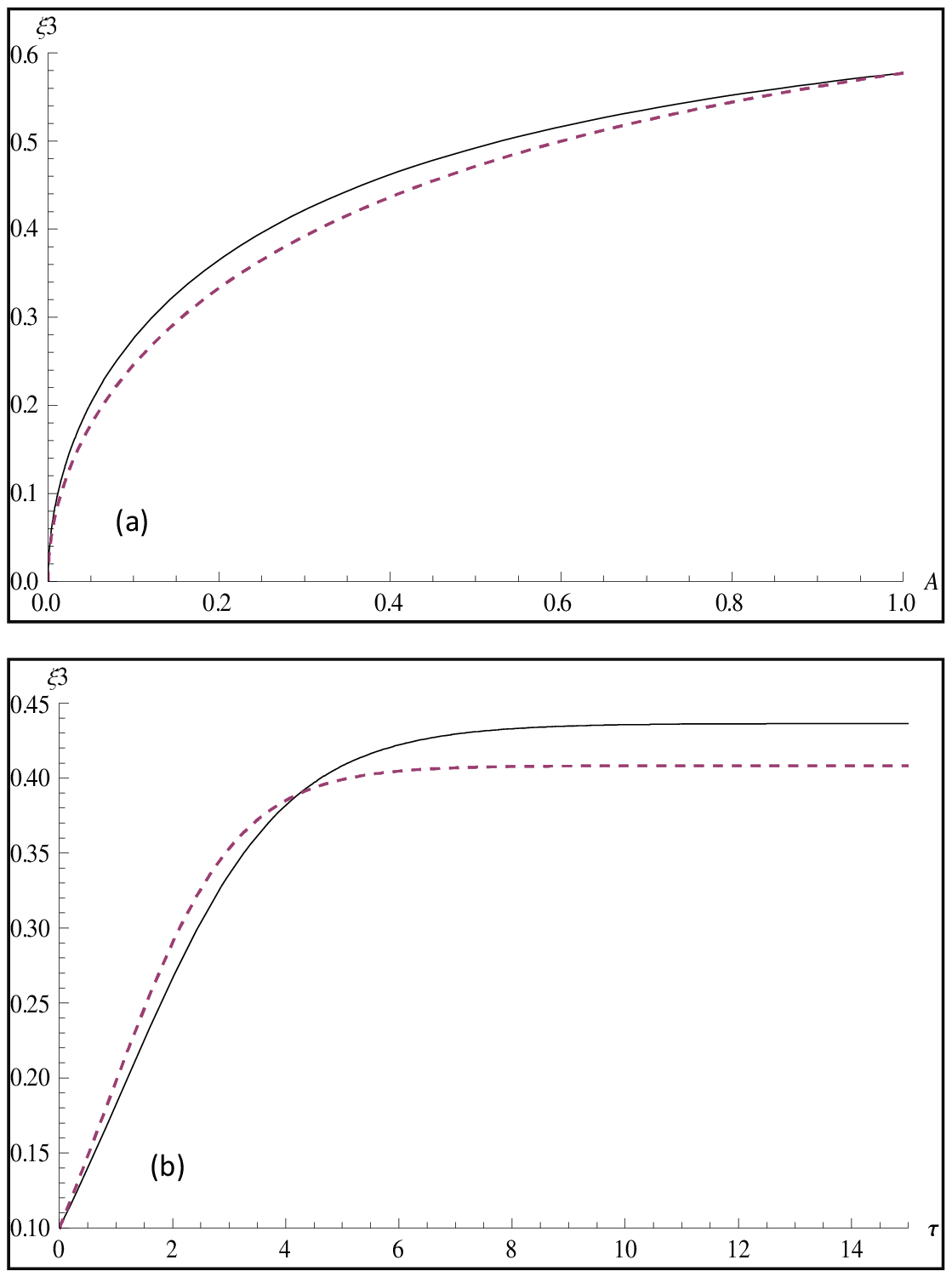}}
\begin{verse}
\vspace{-0.1cm} \caption{(a).Variation of $\xi_3$ against $A$ and
(b).Variation of $\xi_3$ against $\tau$, where the black line
gives the saturation growth rate since by Eqs.(11) and (12)and the
dashed line gives the classical value of growth rate of the tip of
the bubble.}\label{Fig:1}
\end{verse}
\end{figure}

\begin{figure}[p]
\vbox{\hskip 1.cm \epsfxsize=12cm \epsfbox{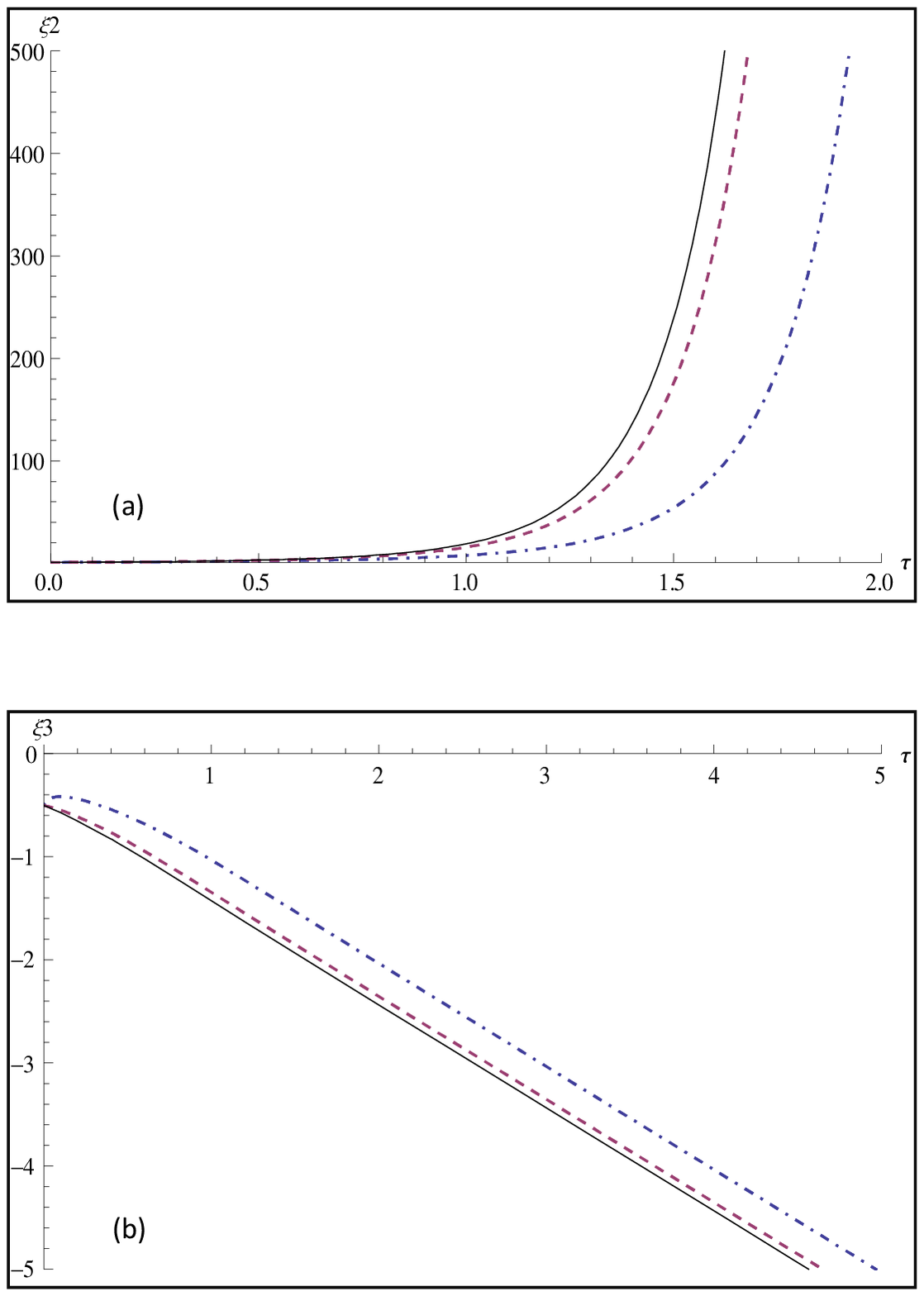}}
\begin{verse}
\vspace{-0.1cm} \caption{(a) Variation of $\xi_2$ against $\tau$
and (b) Variation of $\xi_3$ against $\tau$ with initial value
$\xi_1=-0.1,\xi_2=1.0,\xi_3=-0.5$ and $r$=2(dot-dashed),
5(dashed), 20(black).}\label{Fig.2}
\end{verse}
\end{figure}

\begin{figure}[p]
\vbox{\hskip 1.cm \epsfxsize=12cm \epsfbox{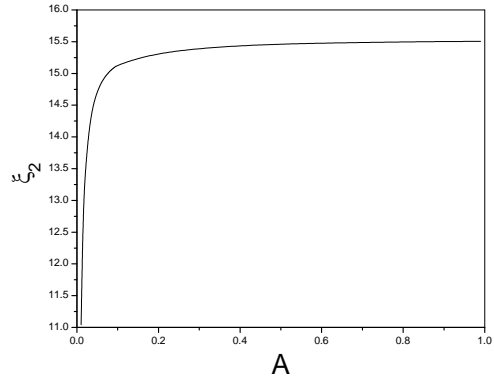}}
\begin{verse}
\vspace{-0.1cm} \caption{Variation of $\xi_{2}$ against $A$ with
initial value $\xi_1=-0.1,\xi_2=1.0,\xi_3=-0.5$.}\label{Fig:3}
\end{verse}
\end{figure}

\begin{figure}[p]
\vbox{\hskip 1.cm \epsfxsize=12cm \epsfbox{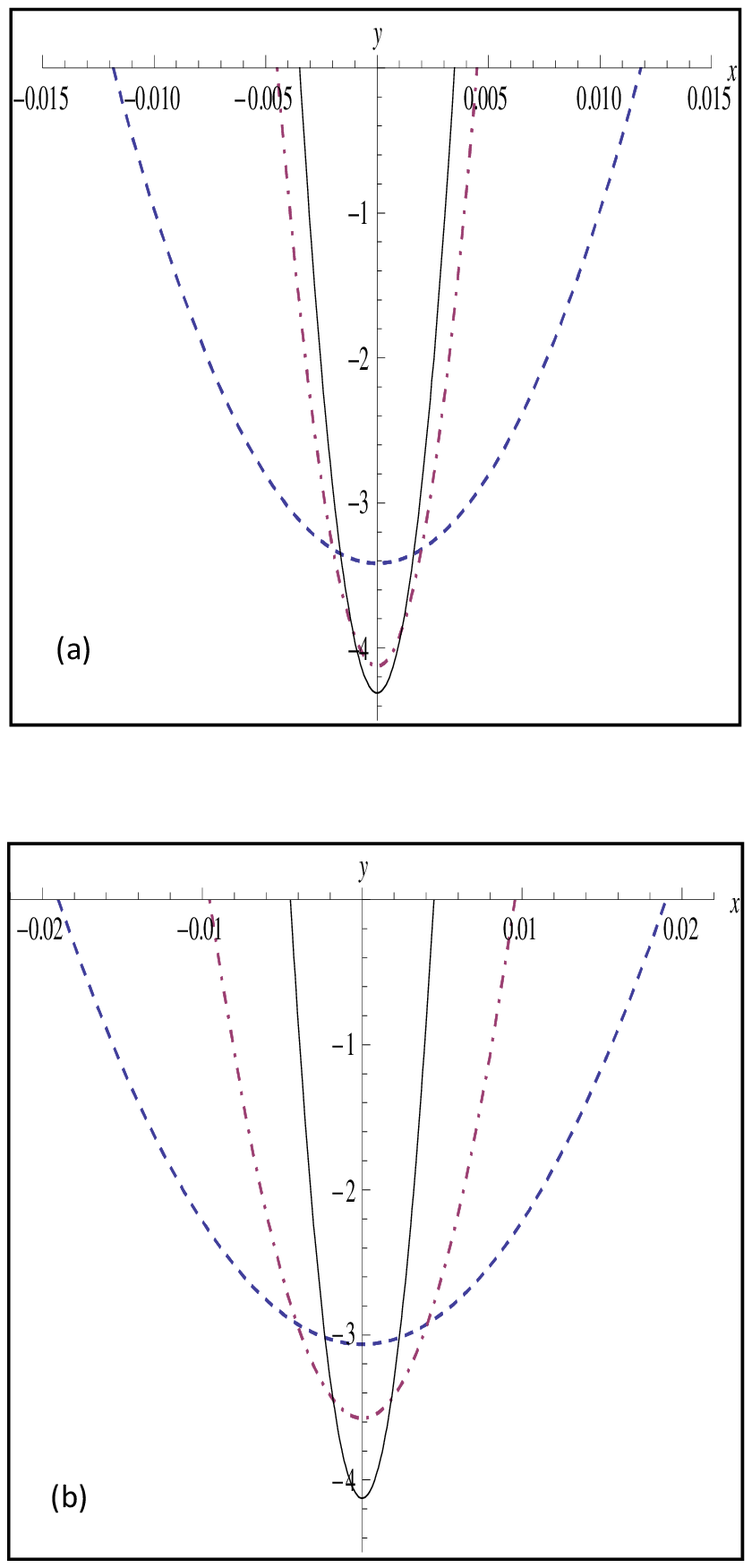}}
\begin{verse}
\vspace{-0.1cm} \caption{Shows formation of spikes for different
values of (a)$A$ = $\frac{1}{3}$(dashed),
$\frac{2}{3}$(dot-dashed), $\frac{19}{21}$(black) for fixed
$\tau$= 2.5 and (b)$\tau$ = 2.1(Dashed), 2.3(dot-dashed),
2.5(black) for fixed $A$ = $\frac{2}{3}$.}\label{Fig:4}
\end{verse}
\end{figure}

\begin{figure}[p]
\vbox{\hskip 1.cm \epsfxsize=12cm \epsfbox{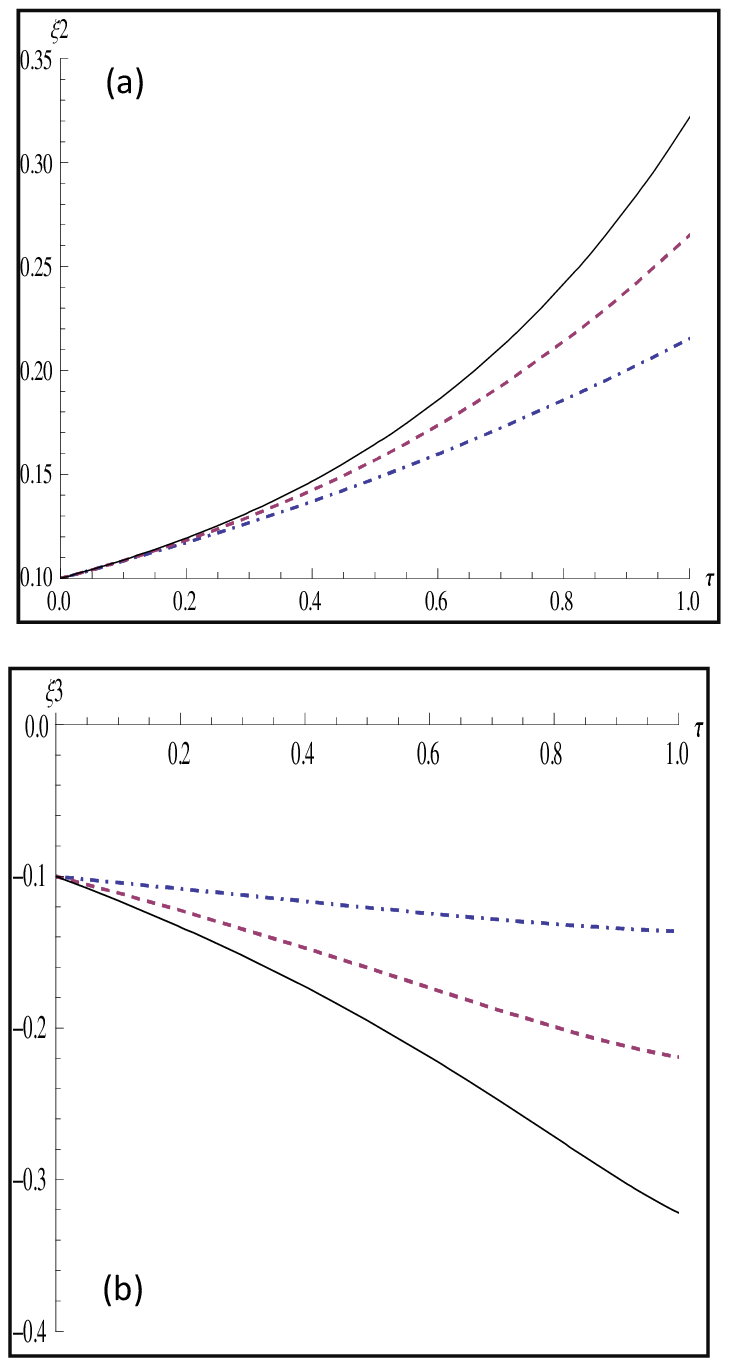}}
\begin{verse}
\vspace{-0.1cm} \caption{(a) Variation of $\xi_2$ against $\tau$
and (b) Variation of $\xi_3$ against $\tau$ with initial value
$\xi_1=-0.1,\xi_2=0.1,\xi_3=-0.1$ and $r$=2(dot-dashed),
5(dashed), 20(black).}\label{Fig.5}
\end{verse}
\end{figure}

\end{document}